\documentclass[aps,pre,twocolumn,groupedaddress,showpacs,showkeys]{revtex4}

\usepackage{graphicx}
\usepackage{dcolumn}
\usepackage{bm}
\usepackage{amssymb}
\usepackage{amsmath}

\begin{document}


\title{SYNCHRONIZATION OF SPECTRAL COMPONENTS \\ AND ITS REGULARITIES IN CHAOTIC
DYNAMICAL SYSTEMS \footnote{Published in Physical Review E 71~(5),
2005, 056204} }


\author{Alexander~E.~Hramov}
\email{aeh@cas.ssu.runnet.ru}
\author{Alexey~A.~Koronovskii}
\email{alkor@cas.ssu.runnet.ru}
\author{Mariya~K.~Kurovskaya}
\author{Olga~I.~Moskalenko}
\affiliation{Faculty of Nonlinear Processes, Saratov State
University, Astrakhanskaya, 83, Saratov, 410012, Russia}




\begin{abstract}
The chaotic synchronization regime in coupled dynamical systems is
considered. It has been shown, that the onset of synchronous
regime is based on the appearance of the phase relation between
interacting chaotic oscillators frequency components of Fourier
spectra. The criterion of synchronization of spectral components
as well as the measure of synchronization have been discussed. The
universal power law has been described. The main results are
illustrated by coupled R\"ossler systems, Van-der-Pol and
Van-der-Pol--Duffing oscillators.
\end{abstract}


\maketitle



\section*{Introduction}
\label{Sct:Intro}

Chaotic synchronization is one of the fundamental phenomena
actively studied last time ~\cite{Pikovsky:2002_SynhroBook},
having important both theoretical and applied significance (e.g.,
for the information transmission by means of deterministic chaotic
signals
\cite{Murali:1993_SignalTransmission,Chua:1997_Criptography}, in
biological~\cite{Elson:1998_NeronSynchro} and
physiological~\cite{Prokhorov:2003_HumanSynchroPRE} tasks, etc.).
Several different types of chaotic synchronization of coupled
oscillators, i.e. generalized synchronization
\cite{Rulkov:1995_GeneralSynchro}, phase synchronization
\cite{Pikovsky:2002_SynhroBook}, lag synchronization
\cite{Rosenblum:1997_LagSynchro} and complete synchronization
\cite{Pecora:1991_ChaosSynchro} are traditionally distinguished.
There are also attempts to find a unifying framework for chaotic
synchronization of coupled
dynamical systems \cite{Brown:2000_ChaosSynchro,%
Boccaletti:2002_SynchroPhysReport,Boccaletti:2001_UnifingSynchro}.

In our works~\cite{Hramov:2004_Chaos,alkor:2004_JETPLetters_TSS}
it was shown that phase, generalized, lag and complete
synchronization are closely connected with each other and, as a
matter of fact, they are different manifestations of one type of
synchronous oscillations behavior of coupled chaotic oscillators
called time scale synchronization. Synchronous regime character
(phase, generalized, lag or complete synchronization) is defined
by the presence of synchronous time scales $s$, introduced by
means of continuous wavelet
transform~\cite{Torresani:1995_WVT,Daubechies:1992_WVTBook,%
alkor:2003_WVTBookEng} with Molrlet mother wavelet function. Each
of time scales can be characterized by the phase $\phi_s(t)=\arg
W(s,t)$, where $W(s,t)$ is the complex wavelet surface. In this
case, the phenomenon of chaotic synchronization of coupled systems
is manifested by a synchronous behavior of the phases of coupled
chaotic oscillators $\phi_{s1,2}(t)$ observed on certain
synchronized time scales range $s_m<s<s_b$, for time scales $s$
from which the phase locking condition
\begin{equation}
|\phi_{s1}(t)-\phi_{s2}(t)|<\mathrm{const}
\label{eq:PhaseLocking}
\end{equation}
is satisfied, and the part of the wavelet spectrum energy being
fallen on this range does not equal zero
(see~\cite{Hramov:2004_Chaos,Hramov:2005_TSS_EWave_Chaos} for
detail). The range of synchronized time scales $[s_m;s_b]$ expands
when the coupling parameter between systems increases. If the
coupling type between oscillators is defined in such a way that
the lag synchronization appearance is possible then all time
scales become synchronized with further coupling parameter
increasing, while the coinciding states of interacting oscillators
are shifted in time relative to each other:
$\mathbf{x}_1(t-\tau)\simeq \mathbf{x}_2(t)$. Further coupling
parameter increasing leads to a decrease of the time shift $\tau$.
The oscillators tend to the regime of complete synchronization,
$\mathbf{x}_1(t)\simeq \mathbf{x}_2(t)$ and the phase difference
$(\phi_{s1}(t)-\phi_{s2}(t))$ tends to be zero on all time scales.

The time scale $s$ introduced into consideration by means of
continuous wavelet transform can be considered as a quantity which
is inversely proportional to the frequency $f$ defined with the
help of Fourier transformation. For the Morlet mother wavelet
function~\cite{Daubechies:1992_WVTBook} with parameter
$\Omega=2\pi$ the relationship between the frequency $f$ and the
time scale is quite simple: $s=1/f$. Therefore, the time scale
synchronization should also manifest in the appearance of the
phase relation between frequency components $f$ of corresponding
Fourier spectra $S(f)$ of interacting oscillators.

In this paper we consider the synchronization of spectral
components of Fourier spectra of coupled oscillators. We discuss
the mechanism  of chaotic synchronization regime manifestation in
coupled dynamical systems based on the appearance of the phase
relation between frequency components of Fourier spectra of
interacting chaotic oscillators (see
also~\cite{Rulkov:1994_PhaseRelation}). One can also consider the
obtained results as a criterion of existence (or otherwise, an
impossibility of existence) of lag synchronization regime in
coupled dynamical systems.

The structure of this paper is the following. In
Section~\ref{Sct:SpectralComponents} we discuss the
synchronization of spectral components of Fourier spectra. We
illustrate our approach with the help of two coupled R\"ossler
systems in Sec.~\ref{Sct:Resslers}. The quantitative measure of
synchronization is described in Section~\ref{Sct:Measure}. The
universal power law taking place in the presence of time scale
synchronization regime is discussed in Sections~\ref{Sct:VdP} and
\ref{Sct:orbits}. The final conclusion is presented in
Section~\ref{Sct:Conclusion}.

\section{Synchronization of spectral components of Fourier spectra}
\label{Sct:SpectralComponents}

It should be noted that the continuous wavelet transform is
characterized by the frequency resolution lower than the Fourier
transformation
(see~\cite{Daubechies:1992_WVTBook,alkor:2003_WVTBookEng}). The
continuous wavelet transform appears as smoothing of the Fourier
spectrum, whereby the dynamics on a time scale $s$ is determined
not only by the spectral component $f=1/s$ of the Fourier
spectrum. This dynamics is also influenced by the neighboring
components as well, the degree of this influence depends both on
their positions in the Fourier spectrum and on their intensities.
Thus, the fact that coupled chaotic oscillators exhibit
synchronization on a time scale $s$ of the wavelet spectrum by no
means implies that the corresponding components $f=1/s$ of the
Fourier spectrum of these systems are also synchronized.

Let $x_1(t)$ and $x_2(t)$ be the time series generated by the
first and the second coupled chaotic oscillators, respectively.
The corresponding Fourier spectra are determined by the relations
\begin{equation}
S_{1,2}(f)=\int\limits_{-\infty}^{+\infty}x_{1,2}(t)e^{-i2\pi f
t}\,dt. \label{eq:FourierTransform}
\end{equation}
Accordingly, each spectral component $f$ of the Fourier spectrum
$S(f)$ can be characterized by an instantaneous phase
$\phi_f(t)=\phi_{f0}+2\pi ft$, where ${\phi_{f0}=\arg S(f)}$.
However, since the phase $\phi_f(t)$ corresponding to the
frequency $f$ of the Fourier spectrum $S(f)$ increases with the
time linearly, the phase difference of the interacting oscillators
at this frequency
${\phi_{f1}(t)-\phi_{f2}(t)}={\phi_{f01}-\phi_{f02}}$ is always
bounded and, hence, the traditional condition of phase entrainment
(used for detection of the phase synchronization regime)
\begin{equation}
|\phi_{1}(t)-\phi_{2}(t)|<\mathrm{const},
\label{eq:PhaseLockingTraditional}
\end{equation}
is useless. Apparently, a different criterion should be used to
detect the synchronization of coupled oscillators at a given
frequency $f$.

In the regime of lag synchronization, the behavior of coupled
oscillators is synchronized on all time scales $s$ of the wavelet
transform (see~\cite{Hramov:2004_Chaos}). Therefore, one can
expect that all frequency components of the Fourier spectra of the
systems under consideration should be synchronized as well. In
this case, $x_1(t-\tau)\simeq x_2(t)$ and, hence, taking into
account (\ref{eq:FourierTransform}) one has to obtain
\begin{equation}
S_2(f)\simeq S_1(f)e^{i2\pi\tau f}.
\label{eq:SpectraRelation}
\end{equation}
Thus, in the case of coupled chaotic oscillators occurring in the
regime of lag synchronization their instantaneous phases
corresponding to the spectral component $f$ of the Fourier spectra
$S_{1,2}(f)$ will be related to each other as
${\phi_{f2}(t)\simeq\phi_{f1}(t)+2\pi\tau f}$ and, hence, the
phase difference $\phi_{f2}(t)-\phi_{f1}(t)$ of coupled
oscillators on the frequency $f$ must obey the relation
\begin{equation}
\begin{split}
\Delta\phi_f&=\phi_{f1}(t)-\phi_{f2}(t)=\\
&=\phi_{f01}-\phi_{f02}=2\pi\tau f. \label{eq:PhaseDifCondition}
\end{split}
\end{equation}
Accordingly, the points corresponding to the phase difference
$\Delta\varphi_f$ of the spectral components of chaotic
oscillators in the regime of lag synchronization on the
$(f,\Delta\phi_f)$ plane must fit a straight line with slope
$k=2\pi\tau$. In the case of complete synchronization of two
coupled identical oscillators the slope of this line $k$ is equal
to zero (see also~\cite{Anishchenko:1992_SynchroOfChaos_IJBC}).

The destroying of the lag synchronization regime (e.g., as a
result of decrease of the coupling strength between oscillators)
and transition to the regime of phase synchronization (in the case
when the instantaneous phase of chaotic signal can be introduced
correctly \cite{Anishchenko:2004_ChaosSynchro}) results in the
loss of synchronism for a part of time scales $s$ of the wavelet
spectra~\cite{Hramov:2004_Chaos}. Accordingly, one can expect that
a part of spectral components of the Fourier spectra in the phase
synchronization regime will also lose synchronism and the points
on the $(f,\Delta\phi_f)$ plane will deviate from the straight
line (\ref{eq:PhaseDifCondition}) mentioned above~\footnote{The
same effect will take place if the instantaneous phase of chaotic
signal can not be introduced correctly due to non-coherent
structure of the chaotic attractor. In this case the phase
synchronization can not be detected, but one can observe the
presence of time scale synchronization.}. It is reasonable to
assume that synchronism will be lost primarily for the spectral
components $f$ characterized by a small fraction of energy in the
Fourier spectra $S_{1,2}(f)$, while the components corresponding
to a greater energy fraction will remain synchronized and the
corresponding points on the $(f,\Delta\phi_f)$ plane will be
located at the straight line as before. As the lag synchronization
regime does not occur in the system any more, the time shift
$\tau$ can be determined by the delay of the most energetic
frequency component $f_m$ in the Fourier spectra
${\tau=(\phi_{f_m2}-\phi_{f_m1})/(2\pi f_m)}$.

As the coupling parameter decreases further, an increasing part of
the spectral components will deviate from synchronism. However, as
long as the most ``energetic'' components remain synchronized, the
coupled systems will exhibit the regime of time scale
synchronization. Obviously, that for the synchronized spectral
component the phase difference $\Delta\varphi_f$ is located after
the transient finished independently on initial conditions.

To describe the synchronization of spectral components, let us
introduce a quantitative characteristic of a number of spectral
components of the Fourier spectra $S_{1,2}(f)$ occurring in the
regime of synchronism,
\begin{widetext}
\begin{equation}
\sigma_L=\\
\frac{\int\limits_{0}^{+\infty}H(|S_1(f)|^2-L)H(|S_2(f)|^2-L)\left(\Delta\phi_{f}-2\pi\tau
f\right)^2\,df}{\int\limits_{0}^{+\infty}H(|S_1(f)|^2-L)H(|S_2(f)|^2-L)\,df},
\label{eq:Sigma_L_int}
\end{equation}
\end{widetext}
where $H(\xi)$ is the Heaviside function, $L$ is the threshold
power level (in dB) above which the spectral components are taken
into account, and $\tau$ is determined by the time shift of the
most energetic frequency component ($f_m$) in the Fourier spectra,
${\tau=(\phi_{f_m2}-\phi_{f_m1})/(2\pi f_m)}$. The quantity
$\sigma_L$ tends to be zero in the regimes of complete and lag
synchronization. After the lag synchronization regime destroying
caused by the decrease of the coupling strength the value of
$\sigma_L$ increases with the number of desynchronized spectral
component of the Fourier spectra $S_{1,2}(f)$ of coupled
oscillators.

Real data are usually represented by discrete time series of a
finite length. In such cases, the continuous Fourier transform
(\ref{eq:FourierTransform}) has to be replaced by its discrete
analog (as it was done
in~\cite{Anishchenko:1992_SynchroOfChaos_IJBC}), and the integral
(\ref{eq:Sigma_L_int}), by the sum
\begin{equation}
\sigma_L=\frac{1}{N}\sum\limits_{j=1}^{N}\left(\Delta\phi_{f_j}-2\pi\tau
f_j\right)^2, \label{eq:Sigma_L}
\end{equation}
taken over all spectral components of the Fourier spectra
$S_{1,2}(f)$ with the power above $L$. In calculating $\sigma_L$,
it is expedient to perform averaging over a set of time series
$x_{1,2}(t)$. The phase shift $\Delta\varphi_f$ can be calculated
either in the way as it was done
in~\cite{Anishchenko:1992_SynchroOfChaos_IJBC} or by means of
cross spectrum~\cite{Noda:2002_CrossSpectraBook}.

\section{Two mutually coupled R\"ossler systems synchronization}
\label{Sct:Resslers}

In order to illustrate the approach proposed above, let us
consider two coupled R\"ossler systems
\begin{equation}
\begin{split}
&\dot
x_{1,2}=-\omega_{1,2}y_{1,2}-z_{1,2}+\varepsilon(x_{2,1}-x_{1,2}),\\
&\dot y_{1,2}=\omega_{1,2}x_{1,2}+ay_{1,2}+\varepsilon(y_{2,1}-y_{1,2}),\\
&\dot z_{1,2}=p+z_{1,2}(x_{1,2}-c), \label{eq:Roessler}
\end{split}
\end{equation}
where $\varepsilon$ is the coupling parameter, ${\omega_1=0.98}$,
${\omega_2=1.03}$. By analogy with the case studied
in~\cite{Rosenblum:2002_FrequencyMeasurement}, the values of
control parameters have been selected as follows: ${a=0.22}$,
${p=0.1}$ and ${c=8.5}$. It is
known~\cite{Rosenblum:2002_FrequencyMeasurement} that two coupled
R\"ossler systems with $\varepsilon=0.05$ occur in the regime of
phase synchronization, while for $\varepsilon=0.15$ the same
systems exhibit lag synchronization.

\begin{figure}[tb]
\centerline{\includegraphics*[scale=0.5]{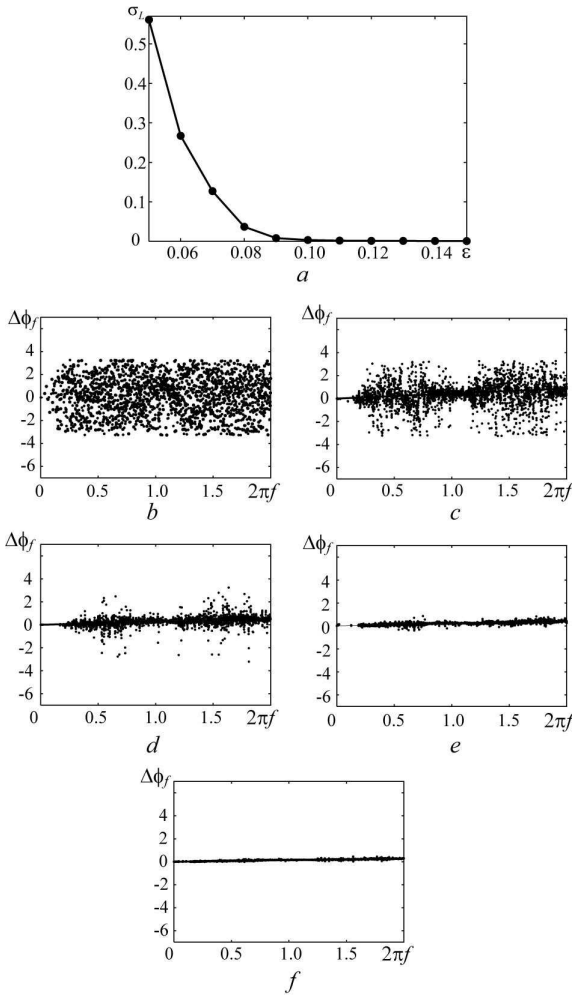}}
\caption{(\textit{a}) The value $\sigma_L$ versus coupling
parameter $\varepsilon$ and (\textit{b})--(\textit{f}) the phase
difference $\Delta\varphi_f$ of various spectral components $f$ of
the Fourier spectra $S_{1,2}(f)$ of two coupled R\"ossler systems
for different values of coupling strength $\varepsilon$.
(\textit{b}) The asynchronous dynamics for coupling parameter
$\varepsilon=0.02$; (\textit{c}) The chaotic synchronization
regime $\varepsilon=0.05$; (\textit{d}), $\varepsilon=0.08$;
(\textit{e}) $\varepsilon=0.1$ and (\textit{f})
$\varepsilon=0.15$. The plots are constructed for the time series
$x_{1,2}(t)$ with a length of 2000 dimensionless time units at a
discretization step of $h=0.2$ at a power level of $L=-40$~dB of
Fourier spectra $S_{1,2}(f)$}
\end{figure}

Figure 1,\textit{a} shows a plot of the value $\sigma_L$ versus
coupling parameter $\varepsilon$. One can see that $\sigma_L$
tends to be zero when the coupling parameter $\varepsilon$
increases, which is the evidence of the transition from phase to
lag synchronization. Figures 1,\textit{b}--1,\textit{f} illustrate
the increase in the number of synchronized spectral components of
the Fourier spectra $S_{1,2}(f)$ of two coupled systems with
coupling strength $\varepsilon$ increasing. Indeed,
Fig.~1,\textit{b} corresponds to the asynchronous dynamics of
coupled oscillators ($\varepsilon=0.02$). There are no synchronous
spectral components for such coupling strength and dots are
scattered randomly over the $(f,\Delta\phi_f)$ plane. The weak
phase synchronization ($\varepsilon=0.05$) after the regime
occurence is shown in the Fig.~1,\textit{c}. There are a few
synchronized spectral components the phase shift $\Delta\varphi_f$
of which satisfies the condition (\ref{eq:PhaseDifCondition}).
Almost all spectral components are non-synchronized, therefore the
points corresponding to the phase differences $\Delta\varphi_f$
are spread over $(f,\Delta\phi_f)$ plane and the value of
$\sigma_L$ is rather large.

\begin{figure}[tb]
\centerline{\includegraphics*[scale=0.4]{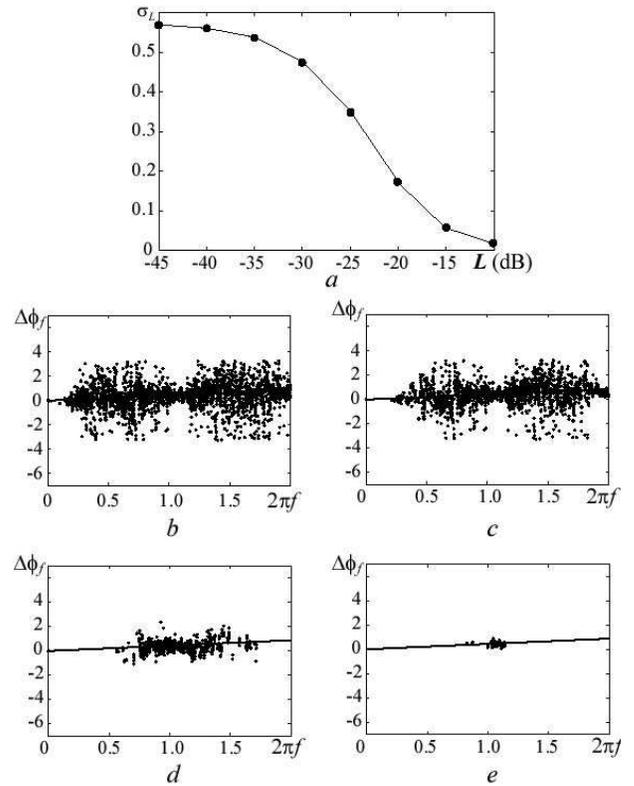}}
\caption{(\textit{a}) The value $\sigma_L$ versus power $L$ at
which the spectral components $f_j$ of the Fourier spectra
$S_{1,2}(f)$ are taken into account in
equation~(\ref{eq:Sigma_L}). (\textit{b})--(\textit{e}) The phase
difference $\Delta\varphi_f$ of various spectral components $f$ of
the Fourier spectra $S_{1,2}(f)$ of two coupled R\"ossler systems
for various power levels $L=-40$ dB (\textit{b}), $L=-30$ dB
(\textit{c}), $L=-20$ dB (\textit{d}) and $L=-10$ dB (\textit{e})
for coupling strength $\varepsilon=0.05$}
\end{figure}

Figures~1,\textit{d} and 1,~\textit{e} correspond to the well
pronounced phase synchronization ($\varepsilon=0.08$ and $0.1$,
respectively). Fig.~1,\textit{f} shows the state of lag
synchronization ($\varepsilon=0.15$), when all spectral components
$f$ of Fourier spectra are synchronized. Accordingly, in this case
all points on the $(f,\Delta\phi_f)$ plane are at the line
(\ref{eq:PhaseDifCondition}) with slope $k=2\pi\tau$. With
coupling strength $\varepsilon$ increasing, the value of
$\sigma_L$ decreases monotonically that verifies the assumption
that when two coupled chaotic systems undergo transition from
asynchronous dynamics to lag synchronization, more and more
spectral components become synchronized. When all spectral
components $f$ are synchronized, the phase shift $\Delta\varphi_f$
for them is $2\pi\tau f$, therefore the points on the
$(f,\Delta\phi_f)$ plane lie on the straight
line~(\ref{eq:PhaseDifCondition}) and the value of $\sigma_L$ is
equal to zero.

Another important question is which spectral components of the
Fourier spectra of interacting chaotic oscillators are
synchronized first and which do it last. Figure~2,\textit{a} shows
a plot of the $\sigma_L$ value for the coupling strength
$\varepsilon=0.05$ (corresponding to the weak phase
synchronization) versus power $L$ at which the spectral components
$f_j$ of the Fourier spectra $S_{1,2}(f)$ are taken into account
in equation~(\ref{eq:Sigma_L}). One can see that the
``truncation'' of the spectral components with small energy leads
to a decrease of the $\sigma_L$ value. Figures
1,\textit{b}--1,\textit{e} illustrate the distribution of the
phase difference $\Delta\varphi_f$ of the spectral components $f$
with the power exceeding the preset level $L$. The data in Fig.~2
show that the most ``energetic'' spectral components are first
synchronized upon the onset of time scale synchronization. On the
contrary, the components with low energies are the first to go out
from synchronism upon destroying of the lag synchronization
regime.

\section{A criterion and measure of synchronization}
\label{Sct:Measure}

Let us briefly discuss a criterion of spectral components
synchronization. Obviously, the relation
(\ref{eq:PhaseDifCondition}) is quite convenient as a criterion of
synchronism in case of lag synchronization destroying in the way
considered above. If the type of coupling between systems has been
defined in such a manner that the lag synchronization regime can
not appear, the relation (\ref{eq:PhaseDifCondition}) can not be
the criterion of spectral components synchronization. So, as a
general criterion of synchronism of identical spectral components
$f$ of coupled systems we have to select the different condition
rather than (\ref{eq:PhaseDifCondition}). As such a criterion we
have chosen the establishment of the phase shift
\begin{equation}
\Delta\varphi_f=\phi_{f01}-\phi_{f02}=\mathrm{const}
\label{eq:PhaseDifCondition_2}
\end{equation}
which must not depend on initial conditions. To illustrate it let
us consider the distribution of phase difference $\Delta\varphi_f$
obtained from the series of $10^3$ experiments for R\"ossler
systems~(\ref{eq:Roessler}). Fig.3,\textit{a} corresponds to the
asynchronous dynamics of coupled oscillators when the coupling
parameter $\varepsilon=0.02$ is below the threshold of chaotic
synchronization appearance (see also Fig.~1,\textit{a}). One can
see, that the phase difference $\Delta\varphi_f$ for the spectral
components $f$ of Fourier spectra $S_{1,2}(t)$ in this case is
distributed randomly over all interval from $-\pi$ to $\pi$. It
means that the phase shift between spectral components $f$ is
different for each experiment (i.e., for different initial
conditions) and, therefore, there is no synchronism whereas the
considered frequency $f$ is the same for both spectra
$S_{1,2}(f)$. The similar distributions are observed for all
spectral components $f$ in the case of asynchronous regime (see
also Fig.~1,\textit{a} and Fig.~3,\textit{b}), though one can
distinguish the maximum in the distribution on the frequency $f$
close to the main frequency of Fourier spectrum $S(f)$ as an
prerequisite of synchronization beginning.

\begin{figure}[tb]
\centerline{\includegraphics*[scale=0.4]{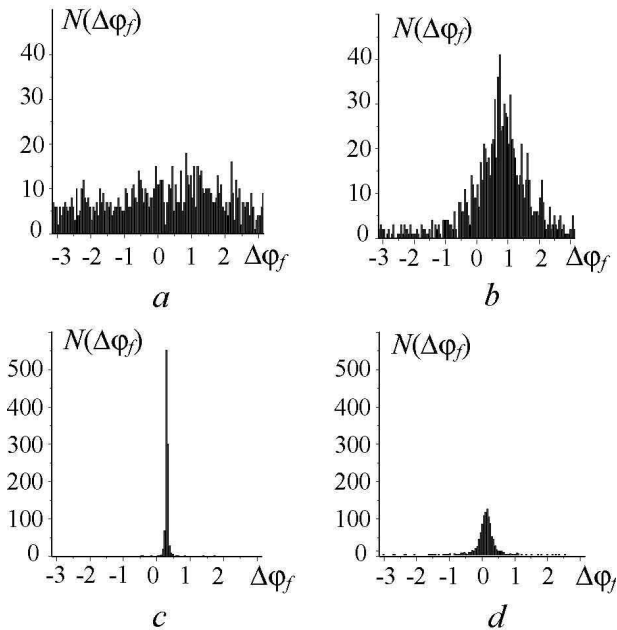}}
\caption{Distribution $N(\Delta\phi_f)$ of phase difference
$\Delta\varphi_f$ obtained from series of $10^3$ experiments  for
R\"ossler systems~(\ref{eq:Roessler}). (\textit{a}) The
asynchronous dynamics takes place ($\varepsilon=0.02$), the
distribution of phase shift $\Delta\varphi_f$ has been obtained
for the spectral components $f\simeq0.0711$; (\textit{b})
$\varepsilon=0.02$, $f\simeq0.1764$; (\textit{c}) the distribution
of phase shift for synchronous spectral component $f\simeq0.1764$
($\varepsilon=0.08$) and (\textit{d}) the analogous distribution
for asynchronous spectral component $f\simeq0.0711$ for the same
coupling parameter $\varepsilon=0.08$. Compare with Fig.~1}
\end{figure}

When the systems demonstrate synchronous behavior the
distributions $N(\Delta\phi_f)$ of phase shift $\Delta\phi_f$ are
quite different. In this case one can distinguish both
synchronized and non-synchronized spectral components
characterized by distributions of phase shift of different types.
In Fig.~3,\textit{c} the distribution of $\Delta\phi_f$ for
synchronous spectral component $\Delta\varphi_f$ is shown. One can
see that it looks like $\delta$-function that means the phase
shift $\Delta\phi_f$ is always the same after transient finished.
Obviously, this phase shift $\Delta\phi_f$ does not depend on
initial conditions.

For the non-synchronized spectral components the distributions
$N(\Delta\phi_f)$ are different (see Fig.~3,\textit{d}).
Evidently, in this case the phase shift $\Delta\phi_f$ is varied
from experiment to experiment. At the same time, the tendency to
the synchronization of these spectral components can be observed.
The distribution $N(\Delta\phi_f)$ looks like Gaussian. With the
coupling parameter increasing the dispersion of it decreases and
the spectral components $f$ of considered Fourier spectra
$S_{1,2}(f)$ tend to be synchronized. The same effect can be
observed in Fig~1,\textit{b-f}. With coupling parameter
$\varepsilon$ increasing, the points on the $(f,\Delta\phi_f)$
plane tend to fit a straight line with slope $k=2\pi\tau$ and
their scattering decreases.

So, the general criterion of synchronism of identical spectral
components $f$ of coupled systems is the establishment of the
phase shift (\ref{eq:PhaseDifCondition_2}) after transient
finished. It is important to note that the case of classical
synchronization of periodical oscillations also obeys to
considered criterion (\ref{eq:PhaseDifCondition_2}) (see, e.g.,
\cite{Adler:1949_Equation}).

Let us consider now the quantitative characteristic of
synchronization. In~\cite{Hramov:2004_Chaos} the measure of
synchronization based on the normalized energy of synchronous time
scales has been introduced. The analogous quantity $\rho$ may be
defined for Fourier spectra $S(f)$ as
\begin{equation}
\rho_{1,2}=\frac{1}{P}\displaystyle\int\limits_{F_{s}}|S_{1,2}(f)|^2
\,df\ \label{eq:rho}
\end{equation}
where $F_s$ is the set of synchronized spectral components and
\begin{equation}
P=\displaystyle\int\limits_0^{+\infty}|S_{1,2}(f)|^2\,df
\end{equation}
is the full energy of chaotic oscillations. In fact, the value of
$\rho$ is the part of the full system energy corresponding to
synchronized Fourier components. This measure $\rho$ is 0 for the
nonsynchronized oscillations and 1 for the case of complete and
lag synchronization regimes as well as the quantity introduced
in~\cite{Hramov:2004_Chaos}. When the systems undergo transition
from asynchronous behavior to the lag synchronization regime the
measure of synchronism takes the value between 0 and 1 that
corresponds to the case when there are both synchronized and
nonsynchronized spectral components in Fourier spectra
$S_{1,2}(f)$.

\begin{figure}[tb]
\centerline{\includegraphics*[scale=0.4]{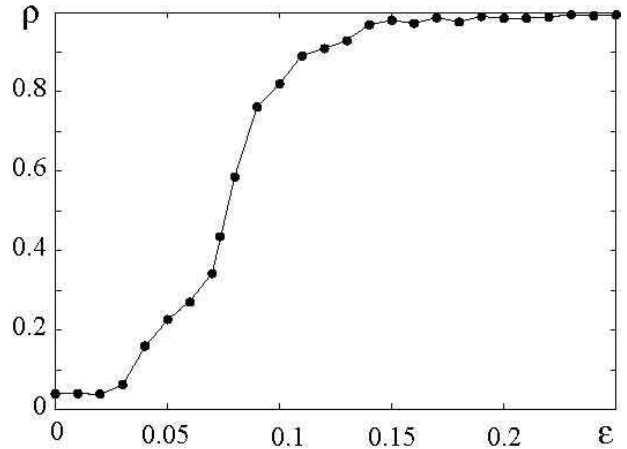}} \caption{The
dependence of the synchronization measure $\rho$ for the first
R\"ossler system (\ref{eq:Roessler}) on coupling parameter
$\varepsilon$}
\end{figure}

For the real data represented by discrete time series of a finite
length one have to use the discrete analog of Fourier transform
while the integrals in the relation~(\ref{eq:rho}) should be
replaced by the sums
\begin{equation}
\rho_{1,2}=\frac{1}{P}\displaystyle\sum\limits_{{j_s}}|S_{1,2}(f_{j_s})|^2\,\Delta
f, \label{eq:rho_sum}
\end{equation}
where
\begin{equation}
P=\displaystyle\sum\limits_{j}\,|S_{1,2}(f_j)|^2\Delta f.
\end{equation}
While the sum in the equation (\ref{eq:rho_sum}) is being
calculated only synchronized spectral components $f_{j_s}$ should
be taken into account.

Figure~4 presents the dependence of the synchronization measure
$\rho$ for the first R\"ossler oscillator of
system~(\ref{eq:Roessler}) on the coupling parameter
$\varepsilon$. It is clear that the part of the energy
corresponding to the synchronized spectral components grows with
the coupling strength increasing.

\section{Spectral components behavior at the presence of synchronization}
\label{Sct:VdP}

Let us now consider how the closed frequency components of two
coupled oscillators behave with increase of the coupling strength
$\varepsilon$. As a model of such situation let us select two
mutually coupled Van-der-Pol oscillators
\begin{equation}
\ddot{x}_{1,2}-\left(\lambda-x^2_{1,2}\right)\dot{x}_{1,2}+
\Omega_{1,2}^2x_{1,2}=\pm \\ \varepsilon(x_{2,1}-x_{1,2}),
\label{eq:VDP}
\end{equation}
where $\Omega_{1,2}=\Omega\pm\Delta$ are slightly mismatched
natural cyclic frequencies, $x_{1,2}$ are the variables,
describing the behavior of the first and the second self-sustained
oscillators, respectively. The parameter $\varepsilon$
characterizes the coupling strength between oscillators.
Nonlinearity parameter $\lambda=0.1$ has been chosen small enough
in order to make oscillations of self--sustained generators close
to the single frequency ones.

Asymmetrical type of coupling in system~(\ref{eq:VDP}) ensures the
appearance of the synchronous regime which is similar to the lag
synchronization in chaotic systems. For such type of coupling the
oscillations in the synchronous regime are characterized by one
frequency $\omega=2\pi f$ while small phase shift
$\Delta\varphi_f$ between time series $x_{1,2}(t)$, decreasing
when coupling strength increases, takes place.

Using the method of complex amplitudes, the solution of
equation~(\ref{eq:VDP}) can be found in the form
\begin{equation}
\begin{split}
&x_{1,2}=A_{1,2}e^{i\omega t}+A_{1,2}^*e^{-i\omega t},\\
&\dot{A}_{1,2}e^{i\omega t}+ \dot{A}^*_{1,2}e^{-i\omega t}=0,
\end{split} \label{eq:VaringAmplitudes}
\end{equation}
where ``*'' means complex conjugation, $\omega$ is the cyclic
frequency, at which oscillations in system (\ref{eq:VDP}) are
realized. One can reduce equations~(\ref{eq:VDP}) and
(\ref{eq:VaringAmplitudes}) to the form
\begin{multline}
\dot{A}_{1,2}=\frac{1}{2}\left(\lambda-|A|^2\right)A+\\
i\frac{1}{2\omega}\left[\left(\Omega_{1,2}^2-\omega^2\right)
A_{1,2}\mp \varepsilon(A_{2,1}-A_{1,2})\right] \label{eq:CompAmp}
\end{multline}
by means of the averaging over the fast changing variables.

Choosing the complex amplitudes in the form of
\begin{equation}
A_{1,2}=r_{1,2}e^{\varphi_{1,2}}, \label{eq:A=R_Phi}
\end{equation}
one can obtain the equations for amplitudes $r_{1,2}$ and phases
$\varphi_{1,2}$ of coupled oscillators as follows:
\begin{equation}
\begin{split}
&\dot{r}_{1,2}=\frac{1}{2}\left(\lambda-|r_{1,2}|^2\right)r_{1,2}
\pm\frac{\varepsilon r_{2,1}}{2\omega}\sin(\varphi_{1,2}-\varphi_{2,1}),\\
\\
&\displaystyle
\dot{\varphi}_{1,2}=\frac{\Omega_{1,2}^2-\omega^2\pm
\varepsilon}{2\omega}\mp\frac{\varepsilon r_{2,1}}{2\omega
r_{1,2}}\cos(\varphi_{1,2}-\varphi_{2,1}). \label{eq:R_Phi}
\end{split}
\end{equation}

The oscillations of two Van-der-Pol generators~(\ref{eq:VDP}) are
synchronized when conditions
\begin{equation}
\dot r_{1,2}=0, \qquad \dot\varphi_{1,2}=0
\end{equation}
are satisfied. Assuming the phase difference of oscillations
$\Delta\varphi=\varphi_2-\varphi_1$ is small enough and taking
into account only components of first infinitesimal order over
$\Delta\varphi$, one can obtain the relation for the phase shift
\begin{equation}
\Delta\varphi_{1,2}=\frac{\lambda\sqrt{\Omega^2+\Delta^2\pm2
\sqrt{\Omega\Delta(\varepsilon+\Omega\Delta)}}}{2\varepsilon+4\Omega\Delta}
\label{eq:DeltaPhi_12}
\end{equation}
and frequency
\begin{equation}
\omega_{1,2}=\sqrt{\Omega^2+\Delta^2\pm2
\sqrt{\Omega\Delta(\varepsilon+\Omega\Delta)}}, \label{eq:omega}
\end{equation}
which correspond to the stable and nonstable fixed points of the
system~(\ref{eq:R_Phi}). From relations ~(\ref{eq:DeltaPhi_12})
and (\ref{eq:omega}) one can see that the phase difference
$\Delta\varphi$ of coupled generators is directly proportional to
the frequency of oscillations $\omega$ and inversely proportional
to the coupling parameter $\varepsilon$ for the small values of
detuning parameter $\Delta$
\begin{equation}
\Delta\varphi\simeq\frac{\lambda\omega}{2\varepsilon}.
\label{eq:DeltaPhi_0}
\end{equation}

So, in the synchronous regime the phase shift $\Delta\varphi$ for
synchronized frequencies obeys the relation
\begin{equation}
\Delta\varphi\sim{\omega}{\varepsilon^{-1}}.
\label{eq:DeltaPhi}
\end{equation}
It is important to note, that the time delay between synchronized
spectral components
\begin{equation}
\tau=\frac{\Delta\varphi}{\omega}\sim \varepsilon^{-1}
\label{eq:tau}
\end{equation}
does not depend upon the frequency, and therefore, the time delays
for all frequencies $f$ are equal to each other. Accordingly, the
phase shift $\Delta\varphi_f$ for the frequency $f$ obeys the
relation~(\ref{eq:PhaseDifCondition}) which is the necessary
conditions for lag synchronization appearance. Evidently, if the
type of coupling between oscillators is selected in such manner
that the phase shift $\Delta\varphi_f$ of synchronized spectral
components satisfies the condition~(\ref{eq:DeltaPhi}), the
appearance of the lag synchronization regime is possible for the
large enough values of the coupling strength. Otherwise, if the
established phase shift does not satisfy the
condition~(\ref{eq:DeltaPhi}), the realization of the lag
synchronization regime in the system is not possible for such kind
of coupling. So, the relation~(\ref{eq:DeltaPhi}) can be
considered as the criterion of the possibility of the existence
(or, otherwise, impossibility of existence) of lag synchronization
regime in coupled dynamical systems.

The regularity~(\ref{eq:tau}) takes place for the large number of
dynamical systems and, probably, is universal. Let us consider the
manifestations of this regularity for several examples of coupled
chaotic dynamical systems.

\begin{figure}[tb]
\centerline{\includegraphics*[scale=0.4]{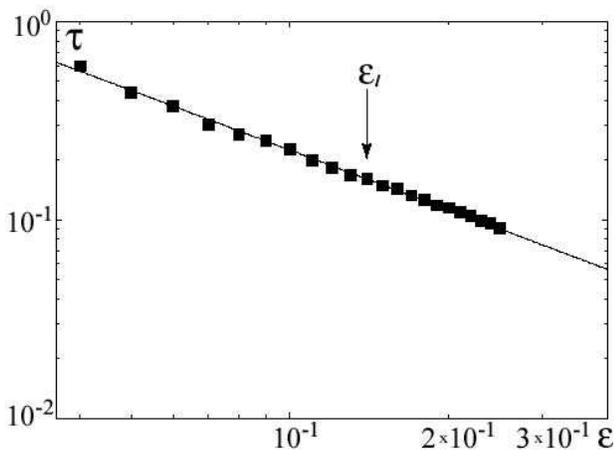}} \caption{The
dependence of time shift $\tau$ between base spectral components
($\blacksquare$ symbols) on coupling parameter $\varepsilon$ for
two coupled R\"ossler systems~(\ref{eq:Roessler}). The straight
line corresponds to the power law $\tau\sim\varepsilon^{-1}$. The
value of coupling parameter $\varepsilon_l\simeq 0.14$
corresponding to the appearance of the lag synchronization regime
is shown by the arrow}
\end{figure}

As the first example we consider the coupled R\"ossler
systems~(\ref{eq:Roessler}) described above. Obviously, one has to
consider the phase shift $\Delta\varphi_f$ (or time shift $\tau$)
of synchronized spectral components to verify the
relation~(\ref{eq:tau}). As  it has been shown above, spectral
components characterized by the large value of energy become
synchronized first when coupling strength increases. So, the main
spectral components $f_{m}$ of Fourier spectra of coupled systems
are synchronized in the most lengthy range of the coupled
parameter value. Therefore, it is appropriate to consider the time
shift $\tau$ of main spectral components for coupling strength
values $\varepsilon>0.05$.

In Fig.~5 the dependence of time lag $\tau$ between
Fourier-spectra base frequency components of interacting chaotic
oscillators on coupling parameter $\varepsilon$ is shown. Base
frequency $\omega_m=2\pi f_m$ of spectrum is close to $\omega=1$
and slightly changes with coupling parameter increasing. In Fig.~5
one can see that after entrainment of Fourier--spectrum base
spectral components of interacting oscillators (that corresponds
to establishment of time scale synchronization regime, see also
\cite{alkor:2004_JETPLetters_TSS}) the time lag $\tau$, which is
between them, obeys the universal power law~(\ref{eq:tau}).

\begin{figure}[b]
\centerline{\includegraphics*[scale=0.4]{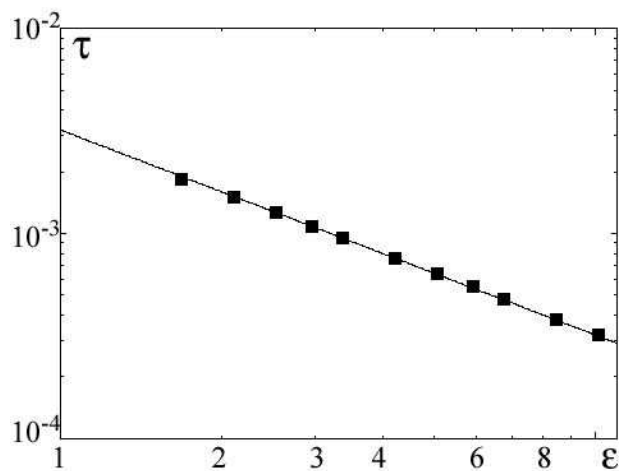}} \caption{The
dependence of time shift $\tau$ between time series $x_1(t)$ and
$x_2(t)$ ($\blacksquare$ symbols) on coupling parameter
$\varepsilon$ for two unidirectionally coupled chaotic
oscillators~(\ref{eq:master}) and (\ref{eq:slave}). The straight
line corresponds to the power law $\tau\sim\varepsilon^{-1}$}
\end{figure}

As the second example we consider the chaotic synchronization of
two unidirectionally coupled Van-der-Pol--Duffing
oscillators~\cite{King:1992_BistableChaosI,%
Murali:1993_SignalTransmission,Murali:1994_SynchroIdenticalSyst}.
The drive generator is described by system of dimensionless
differential equations
\begin{equation}
\begin{split}
&\dot x_1=-\nu_1\left[x_1^3-\alpha x_1 -y_1\right],\\
&\dot y_1=x_1-y_1-z_1,\\ & \dot z_1=\beta y_1,
\end{split}
\label{eq:master}
\end{equation}
while the behavior of the response generator is defined by the
system
\begin{equation}
\begin{split}
&\dot x_2=-\nu_2\left[x_2^3-\alpha x_2 - y_2\right]+\nu_2
\varepsilon (x_1-x_2),\\
& \dot y_2=x_2-y_2-z_2,\\ & \dot z_2=\beta y_2,
\end{split}
\label{eq:slave}
\end{equation}
where $x_{1,2}$, $y_{1,2}$ and $z_{1,2}$ are dynamical variables,
characterizing states of the drive and response generators,
respectively. Values of control parameters are chosen as
following: $\alpha=0.35$, $\beta=300$, $\nu_1=100$, $\nu_2=125$,
the difference of parameters $\nu_1$ and $\nu_2$ provides the
slight nonidentity of considered generators.

In Fig.~6 the dependence of time lag $\tau$ between time
realizations of coupled oscillators on coupling parameter value
$\varepsilon$ is shown. In this range of coupling parameter values
the lag synchronization regime is realized. Obviously, the time
lag $\tau$ also obeys the power law $\tau\sim \varepsilon^n$ with
exponent $n=-1$, that corresponds to the relation~(\ref{eq:tau}).

\section{Unstable periodic orbits}
\label{Sct:orbits}

\begin{figure}[tb]
\centerline{\includegraphics*[scale=0.4]{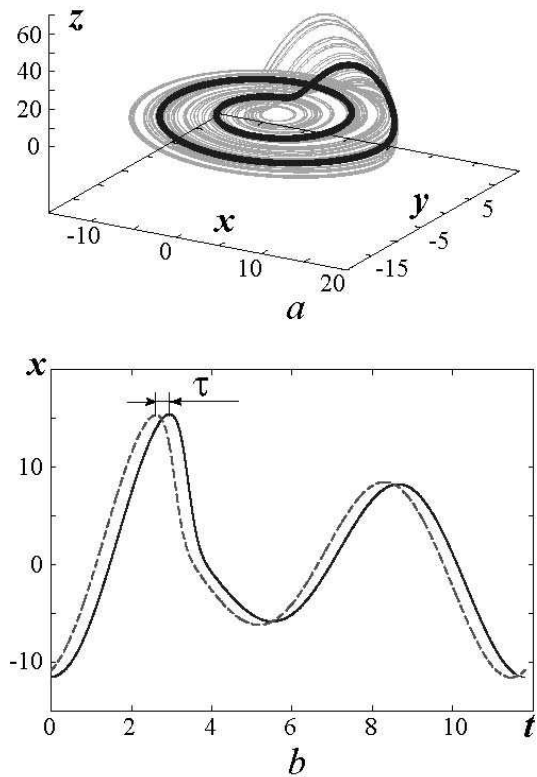}}
\caption{(\textit{a}) The unstable periodic orbit of the length
$m=2$ embedded in the chaotic attractor of the first system.
(\textit{b}) Time series $x_1(t)$ and $x_2(t)$ corresponding to
the ``in--phase'' unstable saddle orbits of length $m=2$ in the
first (solid line) and the second (dashed line) R\"ossler systems,
respectively. The coupling parameter is chosen as
$\varepsilon=0.07$. Time shift $\tau$ is denoted by means of the
arrow}
\end{figure}

It is important to note another manifestation of power
law~(\ref{eq:tau}). It is well known that the unstable periodic
orbits (UPOs) embedded into chaotic attractors play the important
role in the system dynamics~\cite{Cvitanovic:1988_cycles,%
Kostelich:1989_experiment,Cvitanovic:1991_orbits} including the
cases of chaotic synchronization
regimes~\cite{Rulkov:1996_SynchroCircuits,%
Pikovsky:1997_EyeletIntermitt,Pikovsky:1997_PhaseSynchro_UPOs}.
The synchronization of two coupled chaotic systems in terms of
unstable periodic orbits has been discussed in detail in
\cite{Pazo:2002_UPOsSynchro}. It has been shown that UPOs are also
synchronized with each other when the chaotic synchronization in
the coupled systems realizes~\cite{Pazo:2002_UPOsSynchro}. Let us
consider the synchronized saddle orbits $m:n$ ($m=n=1,2,\dots$),
where $m$ and $n$ are the length of unstable periodic orbits of
the first and the second R\"ossler systems~(\ref{eq:Roessler}),
respectively. It was shown that such synchronized orbits may be
both ``in--phase'' and ``out--of--phase'', but only ``in--phase''
orbits exist in all range of coupling parameter values starting
from point of the synchronization beginning
(see~\cite{Pazo:2002_UPOsSynchro} for detail). It is known that
the time shift between synchronized ``in--phase'' orbits decreases
with coupling strength increasing. As the UPOs have an influence
on the system dynamics (and on the Fourier spectra of the
considered systems, too), it seems to be interesting to examine
whether the time shift $\tau$ between UPOs obeys the power
law~(\ref{eq:tau}).

\begin{figure}[tb]
\centerline{\includegraphics*[scale=0.35]{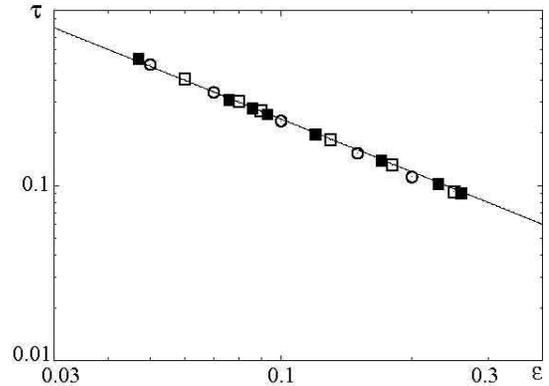}} \caption{The
dependence of time shift $\tau$ between time series $x_1(t)$ and
$x_2(t)$ corresponding to the synchronized saddle orbits of the
first and the second R\"ossler systems on coupling parameter
$\varepsilon$. The symbols $\square$ correspond to unstable orbits
with length $m=1$, the symbols $\bigcirc$ demonstrate the time
shift $\tau$ for the orbits with length $m=2$ and the symbols
$\blacksquare$ show this dependence for orbits with length $m=3$.
The straight line corresponds to the power law
$\tau\sim\varepsilon^{-1}$}
\end{figure}

To calculate the synchronized saddle orbits we have used the
SD--method~\cite{Schmelcher:1998_SDmethodPRE,Pingel:2001_SD-methodPRE}
in the same way as it had been done
in~\cite{Pazo:2002_UPOsSynchro}. The UPO embedded in the chaotic
attractor of the first R\"ossler system and the time series
$x_{1,2}(t)$ corresponding to the ``in--phase'' synchronized UPOs
realized in system~(\ref{eq:Roessler}) for coupling strength
$\varepsilon=0.07$ is shown in Fig.~7. One can see the presence of
the time shift $\tau$ between these time series which can be
easily calculated.

The calculated time shift $\tau$ between such ``in--phase''
synchronized saddle orbits appears to obey the power law
$\tau\sim\varepsilon^{-1}$ as well as the spectral components of
Fourier spectra do (see Fig.~8). We have examined this relation
for ``in--phase'' UPOs with length $m=1\div6$ and found the time
shift dependence on coupling strength agrees with power law
(\ref{eq:tau}) well, but the date is shown in Fig.~8 only for UPOs
with length $m=1\div3$ for the clearness and the simplicity. So,
the power law~(\ref{eq:tau}) seems to be universal and is
manifested in different ways.

\section{Conclusion}
\label{Sct:Conclusion}

In conclusion, we have considered the chaotic synchronization of
coupled oscillators by means of Fourier spectra; several
regularities have been observed.

The chaotic synchronization of coupled oscillators is manifested
in the following way. Starting from the certain coupling parameter
value the synchronization of the main spectral components of
Fourier spectra of interacting chaotic oscillators takes place.
Therefore, for these spectral components $f$ the condition
(\ref{eq:PhaseDifCondition_2}) is satisfied. In this case one can
detect the presence of the time scale synchronization regime
(see~\cite{Hramov:2004_Chaos,Hramov:2005_TSS_EWave_Chaos}). If for
the considered systems one can introduce correctly the
instantaneous phase of chaotic
signal~\cite{Anishchenko:2004_ChaosSynchro,Hramov:2004_Chaos}, one
will also detect easily the phase synchronization by means of
traditional approach (see, e.g., \cite{Pikovsky:2002_SynhroBook}).

With further coupling parameter increasing, more spectral
components become synchronized. If the coupling between
interacting systems is selected in a such way that the lag
synchronization regime can be realized then the time shift between
synchronized spectral components obeys the power
law~(\ref{eq:tau}). The spectral components characterized by the
large value of energy become synchronized first. Accordingly, the
part of energy being fallen on the synchronized spectral
components increases from $0$ (asynchronous dynamics) to 1 (the
lag synchronization regime). Synchronization of all frequency
components corresponds to the appearance of the lag
synchronization regime. With further coupling strength increase,
the time lag $\tau$ obeying the relation~(\ref{eq:tau}) tends to
be zero, and related oscillations tend to demonstrate the complete
synchronization regime. In this case the time shift $\tau$ between
synchronized components does not depend on frequency $f$ of
considered components (it is the same for all synchronized
components) and obeys the power law~(\ref{eq:tau}) with the
exponent $n=-1$. The time shift between synchronized ``in--phase''
UPOs embedded in chaotic attractors also obeys the same power law.

So, in the present paper the mechanism  of chaotic synchronization
regime appearance in coupled dynamical systems, based on the
arising of the phase relation between frequency components of
Fourier spectra of interacting chaotic oscillators has been
discussed. The obtained results concerning the power
law~(\ref{eq:tau}) may be also considered as the criterion of
possible existence (or, otherwise, impossibility of existence) of
lag synchronization regime in coupled dynamical systems (i.e., the
lag synchronization regime can not be observed in the coupled
chaotic oscillators system unless the time shift between
synchronized components obeys the power law ~(\ref{eq:tau})).

\section*{Acknowledgments}
\label{Sct:Acknowledgments}

We thank Michael Zaks for valuable comments and Michael Rosenblum
for critical remarks. We thank also Dr. Svetlana V. Eremina for
English language support. This work was supported by U.S. Civilian
Research and Development Foundation for the Independent States of
the Former Soviet Union (CRDF), grant REC--006), the Supporting
program of leading Russian scientific schools and Russian
Foundation for Basic Research (project 05--02--16273). We also
thank ``Dynastia`` Foundation for financial support.




\begin{thebibliography}{33}
\expandafter\ifx\csname
natexlab\endcsname\relax\def\natexlab#1{#1}\fi
\expandafter\ifx\csname bibnamefont\endcsname\relax
  \def\bibnamefont#1{#1}\fi
\expandafter\ifx\csname bibfnamefont\endcsname\relax
  \def\bibfnamefont#1{#1}\fi
\expandafter\ifx\csname citenamefont\endcsname\relax
  \def\citenamefont#1{#1}\fi
\expandafter\ifx\csname url\endcsname\relax
  \def\url#1{\texttt{#1}}\fi
\expandafter\ifx\csname
urlprefix\endcsname\relax\def\urlprefix{URL }\fi
\providecommand{\bibinfo}[2]{#2}
\providecommand{\eprint}[2][]{\url{#2}}

\bibitem[{\citenamefont{{A. Pikovsky, M. Rosenblum,
 J. Kurths}}(2001)}]{Pikovsky:2002_SynhroBook}
\bibinfo{author}{\bibnamefont{{A. Pikovsky., M. Rosenblum, J. Kurths}}},
  \emph{\bibinfo{title}{Synchronization: a universal concept in nonlinear
  sciences}} (\bibinfo{publisher}{Cambridge University Press},
  \bibinfo{year}{2001}).

\bibitem[{\citenamefont{{K. Murali, M. Lakshmanan}}(1994{\natexlab{a}})}]{Murali:1993_SignalTransmission}
\bibinfo{author}{\bibnamefont{{K. Murali, M. Lakshmanan}}},
  \bibinfo{journal}{Phys. Rev. E} \textbf{\bibinfo{volume}{48}},
  \bibinfo{pages}{R1624} (\bibinfo{year}{1994}{\natexlab{a}}).

\bibitem[{\citenamefont{{T. Yang~T., C.W. Wu and
L.O. Chua}}(1997)}]{Chua:1997_Criptography}
\bibinfo{author}{\bibnamefont{{T. Yang~T., C.W. Wu and L.O. Chua}}},
  \bibinfo{journal}{IEEE Trans. Circuits and Syst.}
  \textbf{\bibinfo{volume}{44}}, \bibinfo{pages}{469} (\bibinfo{year}{1997}).

\bibitem[{\citenamefont{{R.C. Elson, A.I. Selverston, R. Huerta, N.F. Rulkov, M.I. Rabinovich,
H.D.I. Abarbanel}}(1998)}]{Elson:1998_NeronSynchro}
\bibinfo{author}{\bibnamefont{{R.C. Elson, A.I. Selverston, R. Huerta, N.F. Rulkov, M.I. Rabinovich,
H.D.I. Abarbanel}}}, \bibinfo{journal}{Phys.
  Rev. Lett.} \textbf{\bibinfo{volume}{81}}, \bibinfo{pages}{5692}
  (\bibinfo{year}{1998}).

\bibitem[{\citenamefont{{M.D. Prokhorov, V.I. Ponomarenko,
V.I. Gridnev, M.B. Bodrov, A.B.
Bespyatov}}(2003)}]{Prokhorov:2003_HumanSynchroPRE}
\bibinfo{author}{\bibnamefont{{M.D. Prokhorov, V.I. Ponomarenko,
V.I. Gridnev, M.B. Bodrov, A.B. Bespyatov}}},
\bibinfo{journal}{Phys.
  Rev. E} \textbf{\bibinfo{volume}{68}}, \bibinfo{pages}{041913}
  (\bibinfo{year}{2003}).

\bibitem[{\citenamefont{{N.F. Rulkov, M.M. Sushchik, L.S. Tsimring,
H.D.I. Abarbanel}}(1995)}]{Rulkov:1995_GeneralSynchro}
\bibinfo{author}{\bibnamefont{{N.F. Rulkov, M.M. Sushchik, L.S. Tsimring,
H.D.I. Abarbanel}}}, \bibinfo{journal}{Phys. Rev. E}
  \textbf{\bibinfo{volume}{51}}, \bibinfo{pages}{980} (\bibinfo{year}{1995}).

\bibitem[{\citenamefont{{M.G. Rosenblum, A.S. Pikovsky,
 J. Kurths}}(1997)}]{Rosenblum:1997_LagSynchro}
\bibinfo{author}{\bibnamefont{{M.G. Rosenblum, A.S. Pikovsky,
 J. Kurths}}},
  \bibinfo{journal}{Phys. Rev. Lett.} \textbf{\bibinfo{volume}{78}},
  \bibinfo{pages}{4193} (\bibinfo{year}{1997}).

\bibitem[{\citenamefont{{L.M. Pecora,
T.L. Carroll}}(1991)}]{Pecora:1991_ChaosSynchro}
\bibinfo{author}{\bibnamefont{{L.M. Pecora,
T.L. Carroll}}},
  \bibinfo{journal}{Phys. Rev. A} \textbf{\bibinfo{volume}{44}},
  \bibinfo{pages}{2374} (\bibinfo{year}{1991}).

\bibitem[{\citenamefont{{R. Brown,
 L. Kocarev}}(2000)}]{Brown:2000_ChaosSynchro}
\bibinfo{author}{\bibnamefont{{R. Brown, L. Kocarev}}},
  \bibinfo{journal}{Chaos} \textbf{\bibinfo{volume}{10}}, \bibinfo{pages}{344}
  (\bibinfo{year}{2000}).

\bibitem[{\citenamefont{{S. Boccaletti, J. Kurths, G. Osipov, D.L. Valladares,
C.S. Zhou}}(2002)}]{Boccaletti:2002_SynchroPhysReport}
\bibinfo{author}{\bibnamefont{{S. Boccaletti, J. Kurths, G. Osipov, D.L. Valladares,
C.S. Zhou}}}, \bibinfo{journal}{Physics Reports}
  \textbf{\bibinfo{volume}{366}}, \bibinfo{pages}{1} (\bibinfo{year}{2002}).

\bibitem[{\citenamefont{{S. Boccaletti., L.M. Pecora,
A. Pelaez}}(2001)}]{Boccaletti:2001_UnifingSynchro}
\bibinfo{author}{\bibnamefont{{S. Boccaletti., L.M. Pecora,
A. Pelaez}}},
  \bibinfo{journal}{Phys. Rev. E} \textbf{\bibinfo{volume}{63}},
  \bibinfo{pages}{066219} (\bibinfo{year}{2001}).

\bibitem[{\citenamefont{{A.E. Hramov,
A.A. Koronovskii}}(2004)}]{Hramov:2004_Chaos}
\bibinfo{author}{\bibnamefont{{A.E. Hramov,
A.A. Koronovskii}}},
  \bibinfo{journal}{Chaos} \textbf{\bibinfo{volume}{14}}, \bibinfo{pages}{603}
  (\bibinfo{year}{2004}).

\bibitem[{\citenamefont{{A.A. Koronovskii, A.E. Hramov}}(2004)}]{alkor:2004_JETPLetters_TSS}
\bibinfo{author}{\bibnamefont{{A.A. Koronovskii, A.E. Hramov}}},
  \bibinfo{journal}{JETP Letters} \textbf{\bibinfo{volume}{79}},
  \bibinfo{pages}{316} (\bibinfo{year}{2004}).

\bibitem[{\citenamefont{{B. Torresani}}(1995)}]{Torresani:1995_WVT}
\bibinfo{author}{\bibnamefont{{B. Torresani}}},
  \emph{\bibinfo{title}{Continuous wavelet transform}}
  (\bibinfo{publisher}{Paris: Savoire}, \bibinfo{year}{1995}).

\bibitem[{\citenamefont{{A.A. Koronovskii, A.E. Hramov}}(2003)}]{alkor:2003_WVTBookEng}
\bibinfo{author}{\bibnamefont{{A.A. Koronovskii, A.E. Hramov}}},
  \emph{\bibinfo{title}{Continuous wavelet analysis and its applications (In
  Russian)}} (\bibinfo{publisher}{Moscow, Fizmatlit}, \bibinfo{year}{2003}).

\bibitem[{\citenamefont{{I. Daubechies}}(1992)}]{Daubechies:1992_WVTBook}
\bibinfo{author}{\bibnamefont{{I. Daubechies}}}, \emph{\bibinfo{title}{Ten
  lectures on wavelets}} (\bibinfo{publisher}{SIAM}, \bibinfo{year}{1992}).

\bibitem[{\citenamefont{{A.E. Hramov, A.A. Koronovskii, P.V. Popov,
I.S. Rempen}}(2005)}]{Hramov:2005_TSS_EWave_Chaos}
\bibinfo{author}{\bibnamefont{{A.E. Hramov, A.A. Koronovskii, P.V. Popov,
I.S. Rempen}}}, \bibinfo{journal}{Chaos}
\textbf{\bibinfo{volume}{15}},
  \bibinfo{pages}{accepted for publication} (\bibinfo{year}{2005}).


\bibitem[{\citenamefont{{N.F. Rulkov, A.R. Volkovskii, A. Rodriguez-Lozano,
E. Del Rio and M.G.Velarde}}(1994)}]{Rulkov:1994_PhaseRelation}
\bibinfo{author}{\bibnamefont{{N.F. Rulkov, A.R. Volkovskii, A. Rodriguez-Lozano,
E. Del Rio and M.G.Velarde}}}, \bibinfo{journal}{Chaos, Solitons
\& Fractals} \textbf{\bibinfo{volume}{4}}, \bibinfo{pages}{201}
(\bibinfo{year}{1994}).



\bibitem[{\citenamefont{{V.S. Anishchenko, T.E. Vadivasova, D.E. Postnov,
M.A. Safonova}}(1992)}]{Anishchenko:1992_SynchroOfChaos_IJBC}
\bibinfo{author}{\bibnamefont{{V.S. Anishchenko, T.E. Vadivasova, D.E. Postnov,
M.A. Safonova}}}, \bibinfo{journal}{Int. J. Bifurcation and Chaos}
  \textbf{\bibinfo{volume}{2}}, \bibinfo{pages}{633} (\bibinfo{year}{1992}).

\bibitem[{\citenamefont{{V.S. Anishchenko, T.E. Vadivasova}}(2004)}]{Anishchenko:2004_ChaosSynchro}
\bibinfo{author}{\bibnamefont{{V.S. Anishchenko, T.E. Vadivasova}}},
  \bibinfo{journal}{Journal of Communications Technology and Electronics}
  \textbf{\bibinfo{volume}{49}}, \bibinfo{pages}{69} (\bibinfo{year}{2004}).

\bibitem[{\citenamefont{{I. Noda, Y. Ozaki}}(2002)}]{Noda:2002_CrossSpectraBook}
\bibinfo{author}{\bibnamefont{{I. Noda, Y. Ozaki}}},
  \emph{\bibinfo{title}{Two--Dimensional Correlation Spectroscopy: Applications
  in Vibrational and Optical Spectroscopy}} (\bibinfo{publisher}{John Wiley \&
  Sons}, \bibinfo{year}{2002}).

\bibitem[{\citenamefont{{M.G. Rosenblum, A.S. Pikovsky, J. Kurths, G.V. Osipov, I.Z. Kiss, J.L. Hudson}}(2002)}]
{Rosenblum:2002_FrequencyMeasurement}
\bibinfo{author}{\bibnamefont{{M.G. Rosenblum, A.S. Pikovsky, J. Kurths, G.V. Osipov, I.Z. Kiss, J.L. Hudson}}},
  \bibinfo{journal}{Phys. Rev. Lett.} \textbf{\bibinfo{volume}{89}},
  \bibinfo{pages}{264102} (\bibinfo{year}{2002}).

\bibitem[{\citenamefont{{R. Adler}}(1949)}]{Adler:1949_Equation}
\bibinfo{author}{\bibnamefont{{R. Adler}}}, \bibinfo{journal}{Proc. IRE.}
  \textbf{\bibinfo{volume}{34}}, \bibinfo{pages}{351} (\bibinfo{year}{1949}).

\bibitem[{\citenamefont{{G.P. King, S.T. Gaito}}(1992)}]{King:1992_BistableChaosI}
\bibinfo{author}{\bibnamefont{{G.P. King, S.T. Gaito}}},
  \bibinfo{journal}{Phys. Rev. A} \textbf{\bibinfo{volume}{46}},
  \bibinfo{pages}{3092} (\bibinfo{year}{1992}).

\bibitem[{\citenamefont{{K. Murali,
M.
Lakshmanan}}(1994{\natexlab{b}})}]{Murali:1994_SynchroIdenticalSyst}
\bibinfo{author}{\bibnamefont{{K. Murali, M. Lakshmanan}}},
  \bibinfo{journal}{Phys. Rev. E} \textbf{\bibinfo{volume}{49}},
  \bibinfo{pages}{4882} (\bibinfo{year}{1994}{\natexlab{b}}).

\bibitem[{\citenamefont{{P. Cvitanovi\'c}}(1988)}]{Cvitanovic:1988_cycles}
\bibinfo{author}{\bibnamefont{{P. Cvitanovi\'c}}}, \bibinfo{journal}{Phys. Rev.
  Lett.} \textbf{\bibinfo{volume}{61}}, \bibinfo{pages}{2729}
  (\bibinfo{year}{1988}).

\bibitem[{\citenamefont{{D.P. Lathrop,
E.J. Kostelich}}(1989)}]{Kostelich:1989_experiment}
\bibinfo{author}{\bibnamefont{{D.P. Lathrop, E.J. Kostelich}}},
  \bibinfo{journal}{Phys. Rev. A} \textbf{\bibinfo{volume}{40}},
  \bibinfo{pages}{4028} (\bibinfo{year}{1989}).

\bibitem[{\citenamefont{{P. Cvitanovi\'c}}(1991)}]{Cvitanovic:1991_orbits}
\bibinfo{author}{\bibnamefont{{P. Cvitanovi\'c}}}, \bibinfo{journal}{Physica D}
  \textbf{\bibinfo{volume}{51}} (\bibinfo{year}{1991}).

\bibitem[{\citenamefont{{N.F. Rulkov}}(1996)}]{Rulkov:1996_SynchroCircuits}
\bibinfo{author}{\bibnamefont{{N.F. Rulkov}}}, \bibinfo{journal}{Chaos}
  \textbf{\bibinfo{volume}{6}}, \bibinfo{pages}{262} (\bibinfo{year}{1996}).

\bibitem[{\citenamefont{{A. Pikovsky, G. Osipov, M. Rosenblum, M. Zaks,
J. Kurths}}(1997)}]{Pikovsky:1997_EyeletIntermitt}
\bibinfo{author}{\bibnamefont{{A. Pikovsky, G. Osipov, M. Rosenblum, M. Zaks,
J. Kurths}}}, \bibinfo{journal}{Phys. Rev. Lett.}
  \textbf{\bibinfo{volume}{79}}, \bibinfo{pages}{47} (\bibinfo{year}{1997}).

\bibitem[{\citenamefont{{A. Pikovsky, M. Zaks, M. Rosenblum, G. Osipov, J. Kurths}}(1997)}]{Pikovsky:1997_PhaseSynchro_UPOs}
\bibinfo{author}{\bibnamefont{{A. Pikovsky, M. Zaks, M. Rosenblum, G. Osipov, J. Kurths}}},
\bibinfo{journal}{Chaos} \textbf{\bibinfo{volume}{7}},
  \bibinfo{pages}{680} (\bibinfo{year}{1997}).

\bibitem[{\citenamefont{{D. Paz\'o, M. Zaks, J. Kurths}}(2002)}]{Pazo:2002_UPOsSynchro}
\bibinfo{author}{\bibnamefont{{D. Paz\'o, M. Zaks, J. Kurths}}},
  \bibinfo{journal}{Chaos} \textbf{\bibinfo{volume}{13}}, \bibinfo{pages}{309}
  (\bibinfo{year}{2002}).

\bibitem[{\citenamefont{{P. Schmelcher,
F.K. Diakonos}}(1998)}]{Schmelcher:1998_SDmethodPRE}
\bibinfo{author}{\bibnamefont{{P. Schmelcher, F.K. Diakonos}}},
  \bibinfo{journal}{Phys. Rev. E} \textbf{\bibinfo{volume}{57}},
  \bibinfo{pages}{2739} (\bibinfo{year}{1998}).

\bibitem[{\citenamefont{{D. Pingel, P. Schmelcher,
F.K. Diakonos}}(2001)}]{Pingel:2001_SD-methodPRE}
\bibinfo{author}{\bibnamefont{{D. Pingel, P. Schmelcher,
F.K. Diakonos}}},
  \bibinfo{journal}{Phys. Rev. E} \textbf{\bibinfo{volume}{64}},
  \bibinfo{pages}{026214} (\bibinfo{year}{2001}).

\end{thebibliography}

\end{document}